\documentclass[aps,pra,twocolumn,amsmath,amssymb,nofootinbib,superscriptaddress]{revtex4}
\usepackage[english]{babel}
\usepackage{latexsym}
\usepackage{graphics}
\usepackage{epsfig}
\usepackage{color}
\usepackage{bm}
\usepackage{amsmath}
\usepackage{amssymb}

\usepackage[normalem]{ulem}
\usepackage{color}

\newcommand{\be}{\begin{equation}}
\newcommand{\ee}{\end{equation}}

\begin{document}

\title{Energy of $N$ two-dimensional bosons with zero-range interactions}
\author{B.~Bazak}
\affiliation{Institut de Physique Nucl\'eaire, CNRS-IN2P3, Univ. Paris Sud, Universit\'e Paris-Saclay, 91406 Orsay, France}
\affiliation{The Racah Institute of Physics, The Hebrew University, 9190401, Jerusalem, Israel}

\author{D.~S.~Petrov}
\affiliation{LPTMS, CNRS, Univ. Paris Sud, Universit\'e Paris-Saclay, 91405 Orsay, France}

\date{\today}

\begin{abstract}

  We derive an integral equation describing $N$ two-dimensional bosons with zero-range interactions and solve it for the ground state energy $B_N$ by applying a stochastic diffusion Monte Carlo scheme for up to 26 particles. We confirm and go beyond the scaling $B_N\propto 8.567^N$ predicted by Hammer and Son [Phys. Rev. Lett. {\bf 93}, 250408 (2004)] in the large-$N$ limit.

\end{abstract}

%\pacs{34.50.-s, 05.30.Jp, 67.85.-d}

\maketitle

%=====================
\section{INTRODUCTION}
%=====================

Nonrelativistic two-dimensional bosons with zero-range interactions is one 
of the simplest and therefore fundamentally important models in modern
science. It has essentially only one parameter -- the number of particles
$N$. The other parameter, the dimer binding energy $B_2$, can always be set
to unity by rescaling the coordinates. This model neighbors the model of
one-dimensional attractive bosons, which is exactly solvable \cite{McGuire},
and the model of three-dimensional bosons, the zero-range formulation of
which is ill defined since its ground-state energy is not bound from below
because of the Thomas collapse \cite{Thomas}.

Bruch and Tjon \cite{BruTjo79} have shown that no Thomas collapse occurs in
two-dimensions. They have found two bound states (ground and first excited)
of three two-dimensional bosons. The energies of these states have been
calculated over the years by various few-body methods with increasing
accuracy \cite{BruTjo79,AdhDelFre88,NieFedJen97,NieFedJen99,HamSon04,KarMal06,BroKagKla06}. 
In particular, the ground state trimer energy is known to be 
$B_3=16.5226874 B_2$ \cite{HamSon04,KarMal06}.     

Calculations for the four-boson system have been done using finite-range
potentials, but significant range corrections make the convergence to the
zero-range limit problematic \cite{Tjo80,LimNakAka80,VraKil02}.
The current best estimate of the tetramer energy $B_4=197.3(1) B_2$ is
obtained in the zero-range limit by using the effective field theory
\cite{PlaHamMei04}. Note the fast increase of the binding energy with $N$. 

Hammer and Son \cite{HamSon04} have studied $N$-body droplets in the limit 
of large $N$ and found that their energies increase exponentially with $N$,
\be\label{ratio}
B_{N}/B_2\propto r^N=8.567^N,
\ee
and sizes decrease proportionally to $r^{-N/2}=0.3417^N$. This result relies
on the idea of asymptotic freedom and can be obtained in the mean-field
approximation by introducing a logarithmic running of the coupling constant
as a function of the droplet size. The proportionality coefficient in
Eq.~(\ref{ratio}) is currently unknown and there is no theory which
systematically accounts for finite-$N$ corrections starting from the
large-$N$ limit. On the other hand, numerical efforts to approach the
large-$N$ regime from the few-body side have been impeded by the rapid 
growth of the configurational space and the necessity of an extrapolating
procedure to remove finite-range effects. In particular, finite-range
calculations of Blume~\cite{Blu05} carried out for $N \le 7$ were not
conclusive enough to confirm or disprove Eq.~(\ref{ratio}). 
Lee~\cite{Lee06} calculated energies of up to 10 particles by using lattice
effective field theory and observed convergence towards $B_{N}/B_{N-1}=r$
although with a significant errorbar, $r = 8.3(6)$.

In this paper we derive a many-body integral equation to describe
the system directly in the zero-range limit and solve it by using
our recently developed stochastic algorithm \cite{BazPet17}. 
We tabulate $B_N$ for all $N \leq 26$ and claim the values $c_1=-2.06(4)$ and $c_2=-8(2)$ for the first terms in the large-$N$ expansion
\begin{equation}\label{fit}
\ln (B_N r^{-N}/B_2) = c_1 + c_2/N + ...
\end{equation}

%=======================================
\section{Integral equation for N bosons}
%=======================================

Consider the system of $N$ two-dimensional bosons interacting with each other via a zero-range potential characterized by the two-body binding energy $B_2=\kappa_0^2$ (we set $\hbar=m=1$). The $N$-body wave function at the energy $E$ satisfies the Schr\"odinger equation
\begin{equation}\label{Schr}
(-\sum_{i=1}^N \nabla^2_i/2-E)\Psi({\bf r}_1,...,{\bf r}_N)=0
\end{equation}
supplied with the Bethe-Peierls boundary conditions for each pair $i,j$ in the limit ${\bf r}_i\rightarrow {\bf r}_j$ 
\begin{equation}\label{BP}
\Psi \propto \ln (\kappa_0 |{\bf r}_i-{\bf r}_j| e^\gamma /2). 
\end{equation}
The condition (\ref{BP}) substitutes the interaction potential by relating the logarithmically diverging part of the wave function and its regular (constant) part, terms proportional to higher powers of $|{\bf r}_i-{\bf r}_j|$ being omitted.

We then introduce the decomposition 
\begin{widetext}
\begin{equation}\label{Decomp}
\Psi({\bf r}_1,...,{\bf r}_N)=\sum_{i<j}\phi({\bf r}_1,...,{\bf r}_{i-1},{\bf r}_{N-1},{\bf r}_{i+1},...,{\bf r}_{j-1},{\bf r}_N,{\bf r}_{j+1},...,{\bf r}_{N-2};{\bf r}_i,{\bf r}_j),
\end{equation}
where $\phi({\bf r}_1,...,{\bf r}_{N-2};{\bf r}_{N-1},{\bf r}_N)$ satisfies Eq.~(\ref{Schr}), but, in contrast to $\Psi$, is singular only for ${\bf r}_{N-1}\rightarrow {\bf r}_N$. So far we do not impose any particular boundary condition on $\phi$ in this limit, but require that $\phi({\bf r}_1,...,{\bf r}_{N-2};{\bf r}_{N-1},{\bf r}_N)$ be symmetric in its first $N-2$ arguments. For $E<0$ (in this paper we only consider this case) the general form of $\phi$ satisfying the above constraints is (we introduce $M=N-2$) 

\begin{equation}\label{phi}
\phi({\bf r}_1,...,{\bf r}_{M};{\bf r}_{N-1},{\bf r}_N)=\int \frac{d^2 q_1 ... d^2 q_{M}}{(2\pi)^{2M}}F({\bf q}_1,...,{\bf q}_{M}) e^{i\sum_{j=1}^M{\bf q}_j{\bf r}_j-i(\sum_{j=1}^M{\bf q}_j)({\bf r}_{N-1}+{\bf r}_N)/2} \frac{K_0[\kappa({\bf q}_1,...,{\bf q}_{M}) |{\bf r}_{N-1}-{\bf r}_N|]}{2\pi},
\end{equation} 
\end{widetext}
where $F$ is totally symmetric, $K_0$ is the decaying Bessel function, and $\kappa({\bf q}_1,...,{\bf q}_{M})=\sqrt{-E+\sum_{j=1}^M q_j^2/2+(\sum_{j=1}^M{\bf q}_j)^2/4}$. Equation~(\ref{phi}) corresponds to a linear combination of states of $M$ non-interacting bosons with momenta ${\bf q}_1,..., {\bf q}_M$ and a molecule (described by the wave function $K_0$) with the binding energy $\kappa^2$. We have chosen to work in the center-of-mass frame where the molecule momentum equals $-\sum_{j=1}^M{\bf q}_j$. 

The wave function (\ref{Decomp}) with $\phi$ defined by Eq.~(\ref{phi}) is symmetric and satisfies Eq.~(\ref{Schr}) for arbitrary $F$. This freedom is removed by the boundary condition (\ref{BP}) which, because of the symmetry, can be applied for a single pair, say, $N-1,N$. Namely, by considering the limit ${\bf r}_{N}\rightarrow {\bf r}_{N-1}$ we observe that the logarithmic divergence in the sum of Eq.~(\ref{Decomp}) comes only from the component with $i=N-1,j=N$, given explicitly by Eq.~(\ref{phi}). The corresponding logarithmic and regular contributions are obtained by substituting in Eq.~(\ref{phi}) the asymptotic form $K_0(x)=-\ln(xe^\gamma/2)$. All other components in Eq.~(\ref{Decomp}) contribute only to the regular part of $\Psi$ in the limit ${\bf r}_{N-1}\rightarrow {\bf r}_N$. Denoting this latter contribution by ${\cal F}({\bf r}_1,...,{\bf r}_{N-1})$ the boundary condition (\ref{BP}) leads to the equation
\begin{widetext}
\begin{equation}\label{BPexpl}
\int \frac{d^2 q_1 ... d^2 q_M}{(2\pi)^{2M}}F({\bf q}_1,...,{\bf q}_M) e^{i\sum_{j=1}^M{\bf q}_j{\bf r}_j-i(\sum_{j=1}^M{\bf q}_j){\bf r}_{N-1}} \frac{1}{2\pi}\ln\frac{\kappa({\bf q}_1,...,{\bf q}_{M})}{\kappa_0}={\cal F}({\bf r}_1,...,{\bf r}_{N-1}),
\end{equation} 
which, after Fourier transforming, explicitly reads
\begin{align}\label{STM}
&\frac{F({\bf q}_1,...,{\bf q}_M)}{2\pi}\ln\frac{\kappa({\bf q}_1,...,{\bf q}_M)}{\kappa_0}=2\sum_{i=1}^M \int\frac{d^2p_i}{(2\pi)^2}\frac{F({\bf q}_1,...,{\bf q}_{i-1},{\bf p}_i,{\bf q}_{i+1},...,{\bf q_M})}{\kappa^2({\bf q}_1,...,{\bf q}_{i-1},{\bf p}_i,{\bf q}_{i+1},...,{\bf q_M})+({\bf q}_i+\frac{{\bf q}_1+...+{\bf q}_{i-1}+{\bf p}_i+{\bf q}_{i+1}+...+{\bf q_M}}{2})^2} \nonumber \\
&+\sum_{i<j}\int\frac{d^2p_i d^2p_j}{(2\pi)^4}\frac{F({\bf q}_1,...,{\bf q}_{i-1},{\bf p}_i,{\bf q}_{i+1},...,{\bf q}_{j-1},{\bf p}_j,{\bf q}_{j+1},...,{\bf q_M})}{\kappa^2({\bf q}_1,...,{\bf q}_{i-1},{\bf p}_i,{\bf q}_{i+1},...,{\bf q}_{j-1},{\bf p}_j,{\bf q}_{j+1},...,{\bf q_M})+({\bf q}_i-{\bf q}_j)^2/4}(2\pi)^2\delta\left({\bf p}_i+{\bf p}_j+\sum_{k=1}^M{\bf q}_k\right).
\end{align}
\end{widetext}

Equation~(\ref{STM}) is the two-dimensional $N$-body analog of the three-dimensional three-body equation derived by Skorniakov and Ter-Martirosian~\cite{STM}. The formulation of the $N$-body problem based on Eq.~(\ref{STM}) provides a few advantages compared to the one based on Eqs.~(\ref{Schr}) and (\ref{BP}). Very important for us is the fact that Eq.~(\ref{STM}) incorporates the zero-range boundary condition and thus requires no zero-range extrapolation procedure. Another feature is that this formulation reduces the configurational space of the problem by one set of single-particle coordinates giving obvious advantages for small $N$ (3 or 4) where Eq.~(\ref{STM}) can be solved by deterministic methods. However, this point is not essential for us in this work since we aim at significantly larger $N$. In Sec.~\ref{DMC} we develop and apply to Eq.~(\ref{STM}) a stochastic method based on the diffusion Monte Carlo (DMC) technique.

%=====================================
\section{Stochastic method\label{DMC}}
%=====================================

In order to solve Eq.~(\ref{STM}) we adopt the stochastic method developed by us for calculating binding energies and other characteristics of clusters consisting of up to four identical fermions interacting resonantly with another atom of different mass~\cite{BazPet17}. The idea of the method is particularly transparent in the currently discussed bosonic case where the function $F({\bf q}_1,...,{\bf q}_M)$ is symmetric and can be assumed positive for the ground state. The solution procedure is then based on a stochastic process (diffusion of walkers) in the 2$M$-dimensional space $\{{\bf q}_1,...,{\bf q}_M\}$ for which Eq.~(\ref{STM}) is the detailed-balance condition and $F$ is (proportional to) the probability density distribution. We now briefly describe the procedure in the bosonic case (for more technical details applicable in general see Ref.~\cite{BazPet17} and its supplemental material). 

We start with introducing a new function
\begin{equation}\label{f}
f({\bf q}_1,...,{\bf q}_M)=\frac{F({\bf q}_1,...,{\bf q}_M)}{2\pi \prod_{i=1}^M (D+q_i^2)}\ln\frac{\kappa({\bf q}_1,...,{\bf q}_M)}{\kappa_0},
\end{equation}
which will actually be the probability density distribution of walkers. The proportionality factor relating $F$ and $f$ in Eq.~(\ref{f}) is chosen such that $f$ be normalizable, as this is not necessarily the case for $F$. Another requirement for the form of this proportionality factor is to make the sampling and branching (see below) tasks analytical and, therefore, fast. The parameter $D$ shifts the typical momentum of the distribution and influences the convergence rate, but not the final result. Rewriting Eq.~(\ref{STM}) in terms of $f$ we obtain the integral equation
\begin{widetext}
\begin{equation}\label{STMtype}
f=\sum_i\int d^2p_i K({\bf q}_1,...,{\bf q}_M;{\bf p}_i)f({\bf q}_1,...,{\bf p}_i,...,{\bf q}_M)+\sum_{i<j}\int d^2p_i d^2p_j L({\bf q}_1,...,{\bf q}_M;{\bf p}_i,{\bf p}_j)f({\bf q}_1,...,{\bf p}_i,...,{\bf p}_j,...,{\bf q}_M),
\end{equation}
\end{widetext}
where $K$ and $L$ are abbreviations for bulky but straightforward expressions which we do not write explicitly; $K$ and $L$ correspond, respectively, to the first and second integral on the right hand side of Eq.~(\ref{STM}).

We then create an initial distribution of $N_w$ walkers in the space $\{{\bf q}_1,...,{\bf q}_M\}$ and organize an iterative stochastic process of branching and moving them. Namely, at each iteration walkers are subject to two types of operations: 
\begin{itemize}
    \item {\it Single-particle moves:} A walker with coordinates ${\bf q}_1,...,{\bf p}_i,...,{\bf q}_M$ is branched on average $W_i=\int d^2q_i K({\bf q}_1,...,{\bf q}_M;{\bf p}_i)$ times and the $i$-th coordinate of each of the descendants is moved from ${\bf p}_i$ to ${\bf q}_i$ according to the normalized distribution density function $K({\bf q}_1,...,{\bf q}_M;{\bf p}_i)/W_i$. Single-particle moves of this type are similar to the ones we dealt with in the fermionic case~\cite{BazPet17}. Here we have to calculate the normalization integral and sample a product of two two-dimensional Laplacians, $K \propto (D+q_i^2)^{-1}[\kappa^2+({\bf v}+{\bf q}_i)^2]^{-1}$, where ${\bf v}$ is half of the walker's total initial momentum. Both tasks are analytical and very fast. This move is repeated over all walkers and over all coordinates. 
    \item {\it Pair moves:} A walker with coordinates ${\bf q}_1,...,{\bf p}_i,...,{\bf p}_j,...,{\bf q}_M$ is branched  $W_{i,j}=\int d^2q_id^2q_j L({\bf q}_1,...,{\bf q}_M;{\bf p}_i,{\bf p}_j)$ times and coordinates $i$ and $j$ are moved, respectively, from ${\bf p}_i$ and ${\bf p}_j$ to ${\bf q}_i$ and ${\bf q}_j$, where the latter are drawn from $L({\bf q}_1,...,{\bf q}_M;{\bf p}_i,{\bf p}_j)/W_{i,j}$. In fact, due to the delta function in the second term on the right hand side of Eq.~(\ref{STM}) and, therefore, in the function $L$, the value ${\bf q}_i+{\bf q}_j = -{\bf q}_1-...-{\bf p}_i-...-{\bf p}_j-...-{\bf q}_M$ and only the difference ${\bf q}={\bf q}_i-{\bf q}_j$ has to be sampled. Explicitly, the distribution of ${\bf q}$ is governed by a product of three Laplacians, $L\propto [D+({\bf v}+{\bf q})^2]^{-1}[D+({\bf v}-{\bf q})^2]^{-1}(\kappa^2+q^2)^{-1}$. The normalization integral is analytical and one can find a very fast rejection-based sampling procedure. This move is repeated over all walkers and over all pairs of coordinates.
\end{itemize}
We emphasize that fast branching and sampling is essential for the efficient implementation of this method for large $N\sim M$, particularly, since the number of pair moves grows proportionally to $M^2$. In this context we note that the problem of several ideal bosons resonantly interacting with an impurity is technically easier as there are no pair terms in the corresponding integral equation.

Up to the noise related to the probabilistic branching procedure the walker population at a next iteration is given by $\sum_{i=1}^{M}W_i+\sum_{i<j} W_{i,j}$ summed over all walkers. If this number is systematically larger or smaller than the population at the previous iteration, we can correct the value of $E$, which implicitly enters in the weights $W$ through the kernels $K$ and $L$. Alternatively, and this is what we do in practice, we fix $E=-1$ and use $\kappa_0$ as the parameter that controls the walker population and keeps it close to the initial $N_w$. More precisely, we adjust this parameter during every iteration in such a way that the sum of all branching weights of all walkers equals $N_w$. 

Assuming that a steady equilibrium distribution of walkers is reached and that $\kappa_0$ is fixed at its exact value, it is easy to show that the detailed-balance equation for the above iterative scheme is exactly Eq.~(\ref{STMtype}). With the population correction implemented, we do reach a steady state, but $\kappa_0$ fluctuates around a certain average value $\kappa_0(N_w)$. The amplitude of these fluctuations decreases with $N_w$ and $\kappa_0(N_w)$ converges to the exact value in the limit $N_w\rightarrow \infty$ (see Sec.~\ref{Sec:Results}). 

%===================================
\section{RESULTS\label{Sec:Results}}
%===================================

The free parameters of our algorithm are $D$ and $N_w$. We find
phenomenologically that $2\lesssim D\lesssim 6$ maximizes the convergence
rate. Results presented below are obtained with $D={4}$.  

\begin{figure}
\begin{center}
\includegraphics[width=8.6 cm]{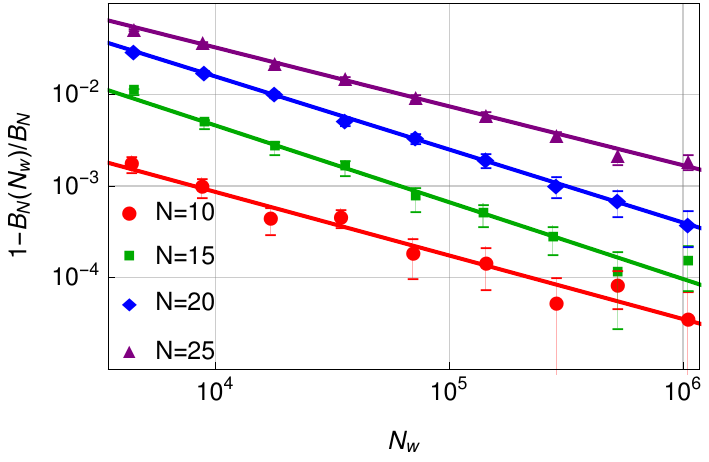}
\caption{\label{fig:Convergence}
The quantity $1-B_N(N_w)/B_N$ versus $N_w$ demonstrating the convergence of our results in the large-$N_w$ limit. $B_N$ is determined from fitting the data with $B_N(N_w)=B_N+cN_w^{-\gamma}$. The straight lines are the corresponding fits.}
\end{center}
\end{figure}

For a given walker population $N_w$, after thermalization and accumulation of statistics, our algorithm gives the quantity $\kappa_0(N_w)$ at fixed $E=-1$ with a certain statistical uncertainty. We then translate it into the $N$-body energy estimate $B_N(N_w)/B_2=1/\kappa_0^2(N_w)$. 
In order to verify the convergence of $B_N(N_w)$ towards $B_N$ for a given $N$ we run our code with various values of $N_w$ and fit the obtained function $B_N(N_w)$ to the form $B_N(N_w)=B_N+cN_w^{-\gamma}$, where $B_N$, $c$, and $\gamma$ are fitting parameters. The best-fit values of the exponent $\gamma$ for different $N$ all belong to the interval $(0.5,1)$. 
In Fig.~\ref{fig:Convergence} we show the quantity $1-B_N(N_w)/B_N$ for a few representative values of $N$ together with the corresponding fits. The energies obtained by using this extrapolating procedure are reported in Tab.~\ref{tbl:Energies}. The error bars represent the sum of the statistical uncertainty and systematic uncertainty which we define as $|B_N(N_{w,{\rm max}})-B_N|$, i.e., the deviation of the energy calculated with the largest walker pool from the extrapolated value. We should note here that the intrinsic time scale of our stochastic process is related only to the dimension of the configurational space and we benefit from relatively short thermalization and correlation times, at most a few hundreds of iterations for largest $N$. Thus, at approximately the same CPU cost we have a large room for increasing $N_w$ and decreasing the number of iterations, which minimizes the finite-$N_w$ systematic error while keeping the statistical uncertainty constant.

\begin{table}
\begin{center}
  \caption{The ground-state energy of the
    $N$ boson droplet in units of the dimer energy. Previous calculations for $N=3,4$ give $B_3/B_2=1.6522688(1) \times 10^1$ \cite{HamSon04} and $B_4/B_2=1.973(1) \times 10^2$ \cite{PlaHamMei04}.
    \label{tbl:Energies}}
\vspace{0.3cm}
{\renewcommand{\arraystretch}{1.25}%
\begin{tabular}
{c@{\hspace{3mm}}
c@{\hspace{8mm}}  c@{\hspace{3mm}} c}
\hline\hline
$N$ & $B_N/B_2$ & $N$ & $B_N/B_2$ \\
\hline
3  & 1.65225(2)$\times 10^1$   & 15 & 8.135(2)$\times 10^{12}$\\ 
4  & 1.9720(1) $\times 10^2$   & 16 & 7.129(4)$\times 10^{13}$\\
5  & 2.0745(1) $\times 10^3$   & 17 & 6.232(2)$\times 10^{14}$\\
6  & 2.0471(1) $\times 10^4$   & 18 & 5.438(3)$\times 10^{15}$\\
7  & 1.9462(1) $\times 10^5$   & 19 & 4.734(2)$\times 10^{16}$\\
8  & 1.8070(1) $\times 10^6$   & 20 & 4.119(2)$\times 10^{17}$\\
9  & 1.6508(4) $\times 10^7$   & 21 & 3.577(2)$\times 10^{18}$\\
10 & 1.4905(2) $\times 10^8$   & 22 & 3.108(4)$\times 10^{19}$\\
11 & 1.3345(2) $\times 10^9$   & 23 & 2.694(5)$\times 10^{20}$\\
12 & 1.1873(4) $\times 10^{10}$& 24 & 2.332(4)$\times 10^{21}$\\
13 & 1.0508(3) $\times 10^{11}$& 25 & 2.018(4)$\times 10^{22}$\\
14 & 9.2596(9) $\times 10^{11}$& 26 & 1.748(4)$\times 10^{23}$\\
\hline\hline
\end{tabular}}
\end{center}
\end{table}

We can benchmark our results for $N=3$ and 4 against the most accurate
previous calculations. 
For the trimer we have $B_3/B_2=16.5225(2)$, in good agreement with
$B_3/B_2=16.522688(1)$ \cite{HamSon04} and $B_3/B_2=16.5226874$
\cite{KarMal06}. 
Our tetramer result $B_4/B_2=197.20(1)$ is more precise than the previously
known value $B_4/B_2=197.3(1)$ \cite{PlaHamMei04}.

\begin{figure}
\begin{center}
\includegraphics[width=8.6 cm]{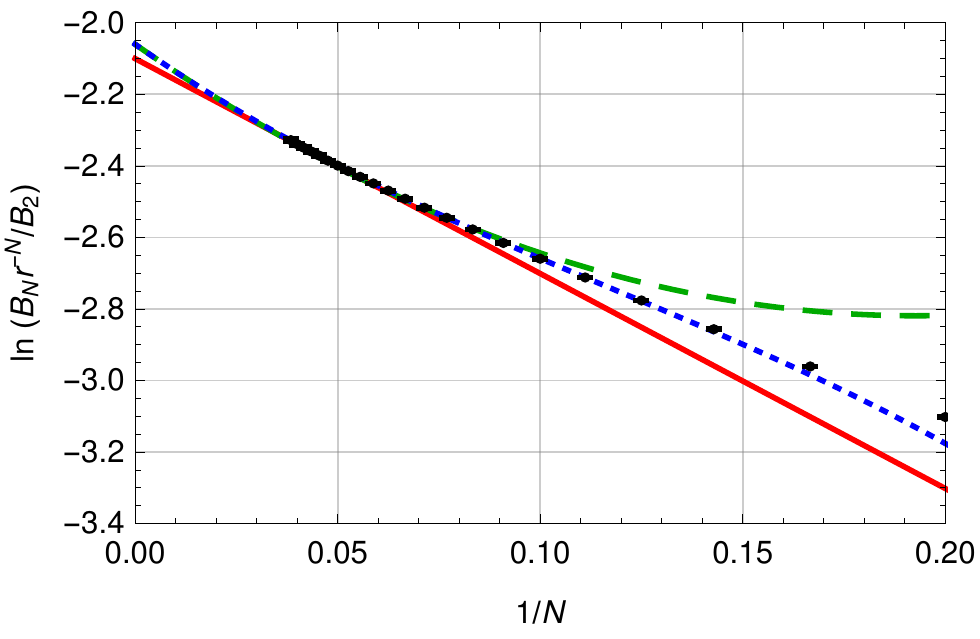}
\caption{\label{fig:Energies}
  $\ln\left(B_N r^{-N}/B_2\right)$ versus $1/N$ (symbols) and linear (solid), quadratic (dashed), and cubic (dotted) fits to the data (see text).
  }
\end{center}
\end{figure}

Our data for larger $N$ are consistent with the value $r=8.567$ \cite{HamSon04,note} and with the assumption that the function $\ln\left(B_N r^{-N}/B_2\right)$ can be expanded in powers of $1/N$. In Fig.~\ref{fig:Energies} we plot this function together with linear, quadratic, and cubic polynomial fits $\sum_i c_i N^{1-i}$ based, respectively, on the data for $N\geq 22$, $N\geq 16$, and $N\geq 10$. The fitting parameters $\{c_1,c_2,...\}$\cite{note2} are, respectively, $\{-2.1, -6.01\}$,  $\{-2.06, -7.88, 20.45\}$, and $\{-2.06, -7.94, 27.2, -77\}$. Based on these findings we claim $c_1=-2.06(4)$ and $c_2=-8(2)$. 

%============================
\section{Summary and outlook}
%============================

In this paper we have calculated the energies of two-dimensional bosonic droplets with strictly zero-range interactions for $N\leq 26$. Our results are consistent with the proportionality factor $r=8.567$ in Eq.~(\ref{ratio}) and we estimate the leading finite-$N$ corrections; we claim $B_N/B_2\approx r^N\exp(c_1+c_2/N)$ where $c_1=-2.06(4)$ and $c_2=-8(2)$. That $c_2$ turns out to be rather large indicates significant finite-$N$ effects for $N$ as large as 10. Our results should be useful as a reference point for developing a theory beyond Ref.~\cite{HamSon04} expected, in particular, to describe the droplet dynamics and excitations. The calculation of excited states with our current method is challenging since the solution in this case is a sign-changing function. 

The exponential scaling of the size and energy raises obvious questions concerning the feasibility of observing such droplets. In the quasi-two-dimensional geometry for small ratio of the scattering length $a<0$ to the confinement oscillator length $l$ the dimer size is exponentially large, $\propto l\exp(\sqrt{\pi/2} l/|a|)$ \cite{PetShl2001}. The $N$-body droplet size, in this case proportional to $l\exp(\sqrt{\pi/2} l/|a|-N\ln\sqrt{r})$, should be much larger than $l$ and smaller than the waist of the laser beams used for the quasi-two-dimensional confinement. This defines the window of parameters where the droplet is observable; one can see that $N$ is not necessarily very small. Given extremely low values of $a$ routinely produced in experiments (see, for example, \cite{Fattori,Hulet,Bourdel}), it is realistic to observe droplets of a few tens of particles. The preparation of a droplet can be realized, for example, by simultaneously switching off of the radial confinement and sweeping the scattering length from a positive to negative value near a zero crossing.

Our method can be generalized to include the leading-order effective-range correction to the $N$-body energy (as has been done in three dimensions \cite{BazPet17}). One can then estimate the energy correction due to finite-range effects associated with the finite width of the cloud in the quasi-two-dimensional case. In addition, it should provide a link to few-body calculations based on realistic finite-range potentials. Finally, we mention that our method can be extended to bosonic mixtures with different masses.

%========================
\section*{ACKNOWLEDGMENT}
%========================
We thank G.~Astrakharchik, J.~Carlson, and D.~Lee for useful communications
and O. Mustaki for technical assistance.
This research was supported by the European Research Council 
(FR7/2007-2013 Grant Agreement No. 341197) and 
from the Pazi Fund. We also acknowledge support by the IFRAF Institute.


\begin{thebibliography}{99}
\bibitem{note}
  We agree with $r=8.567$ within the uncertainty of $0.3\%$, which equals the
  relative error of $B_{N}/B_{N-1}$ for a few largest values of $N$ in Table I.
\bibitem{note2}
  Here $r=8.567$ is not a fitting parameter. By fixing $r$ we determine $c_i$'s
  with a significantly higher precision.
\end{thebibliography}
\end{document}